\documentclass[
 reprint,
superscriptaddress,
nofootinbib,
 amsmath,amssymb,
 aps,
]{revtex4-2}

\usepackage{graphicx}
\usepackage{dcolumn}
\usepackage{bm}
\usepackage{hyperref}
\hypersetup{
    colorlinks=true,
    allcolors=blue
}
\usepackage{xcolor}
\usepackage{dsfont}

\usepackage{tabu}

\newcommand{\ket}[1]{| #1\rangle}

\begin{document}

\title{High-Fidelity Entangled States in a Connectivity-Four Fluxonium Quantum Processor}

\author{J.~Schirk}
\email{johannes.schirk@wmi.badw.de}
\affiliation{Technical University of Munich, TUM School of Natural Sciences, Department of Physics, 85748 Garching, Germany}
\affiliation{Walther-Meißner-Institut, Bayerische Akademie der Wissenschaften, 85748 Garching, Germany}
\author{N.~Bruckmoser}
\author{S. M.~Taubenberger}
\author{F.~Wallner}
\author{N. J.~Glaser}
\author{M.~Zetzl}
\author{L.~Huang}
\author{I.~Tsitsilin}

\affiliation{Technical University of Munich, TUM School of Natural Sciences, Department of Physics, 85748 Garching, Germany}
\affiliation{Walther-Meißner-Institut, Bayerische Akademie der Wissenschaften, 85748 Garching, Germany}
\author{M.~Werninghaus}
\author{L.~Södergren}

\author{K.~Liegener}
\author{C. M. F.~Schneider}
\affiliation{Technical University of Munich, TUM School of Natural Sciences, Department of Physics, 85748 Garching, Germany}
\affiliation{Walther-Meißner-Institut, Bayerische Akademie der Wissenschaften, 85748 Garching, Germany}
\author{S.~Filipp}
\affiliation{Technical University of Munich, TUM School of Natural Sciences, Department of Physics, 85748 Garching, Germany}
\affiliation{Walther-Meißner-Institut, Bayerische Akademie der Wissenschaften, 85748 Garching, Germany}
\affiliation{Munich Center for Quantum Science and Technology (MCQST), 80799 München, Germany}
\affiliation{Munich Quantum Valley GmbH, 85748 Garching, Germany}
\date{\today}

\begin{abstract}
A central challenge in fluxonium-based quantum processors is the extension of the qubit connectivity to two-dimensional lattices compatible with quantum code-error correction. 
Here, we present a fluxonium quantum processor that employs lumped-element resonator couplers which realizes, for the first time, a connectivity-four unit cell with suppressed parasitic interactions.
We achieve parallel single-qubit gate fidelities exceeding 99.9\,\% in simultaneous randomized benchmarking experiments, while maintaining residual static ZZ interactions below 1 kHz across all coupled qubit pairs. 
We implement resonator-induced phase (RIP) gates and benchmark two-qubit gate fidelities exceeding 99\,\% using interleaved randomized benchmarking. To cancel spectator errors observed in two-qubit operations, we implement a refocused RIP gate, recovering coherent control in the presence of multi-qubit connectivity.  
Furthermore, we prepare Greenberger-Horne-Zeilinger states of up to five qubits with a tomographic fidelity of 90\,\%, verifying multi-qubit entanglement within the unit cell. 
These results establish the fluxonium-resonator-fluxonium architecture as a viable approach to realizing densely connected fluxonium processors and provide a scalable path toward quantum error-correction-compatible processor architectures.
\end{abstract}

\maketitle

\section{\label{section:introduction}Introduction}
In the rapidly developing field of quantum computing, superconducting qubits have emerged as a leading platform for realizing large-scale quantum processors~\cite{krantzQuantumEngineersGuide2019, blaisCircuitQuantumElectrodynamics2021}.
Continuous improvements in qubit coherence times, gate fidelities, and fabrication techniques have established superconducting quantum circuits as one of the most mature and scalable hardware architectures for quantum information processing~\cite{mckayBenchmarkingQuantumProcessor2023,acharyaQuantumErrorCorrection2025, blandMillisecondLifetimesCoherence2025}.
In particular, quantum error correction protocols such as the surface code~\cite{dennisTopologicalQuantumMemory2002, fowlerSurfaceCodesPractical2012} impose stringent requirements on both the quality and connectivity of physical qubits: individual gate fidelities must surpass fault-tolerance thresholds, and each qubit must couple to at least four nearest neighbors to populate the two-dimensional lattice that the surface code demands~\cite{litinskiGameSurfaceCodes2019, beverlandSurfaceCodeCompilation2022}.
Meeting these requirements while suppressing the residual interactions that inevitably arise from multi-qubit connectivity remains a central challenge for scaling up superconducting quantum processors.

Among the different superconducting qubit implementations, the fluxonium qubit has attracted increasing attention as a promising alternative to the widely deployed transmon~\cite{manucharyanFluxoniumSingleCooper2009, nguyenHighCoherenceFluxoniumQubit2019a}.
Owing to its large inductive shunt, the fluxonium exhibits an insensitivity to offset charges, a deeply anharmonic energy spectrum and competitive relaxation and coherence times~\cite{somoroffMillisecondCoherenceSuperconducting2021,dingHighFidelityFrequencyFlexibleTwoQubit2023a, zhangUniversalFastFluxControl2021}.
These properties enable single-qubit gate fidelities above 99.99\,\%~\cite{dingHighFidelityFrequencyFlexibleTwoQubit2023a,rowerSuppressingCounterRotatingErrors2024} and are expected to lower the overhead required for fault-tolerant computation through an overall lower error rate~\cite{nguyenBlueprintHighPerformanceFluxonium2022a}.
Despite these intrinsic advantages, scaling fluxonium-based processors beyond linear or minimal-connectivity configurations remains an open engineering problem, limited by the lack of a coupling scheme that simultaneously suppresses residual qubit-qubit interactions during idling and while executing single-qubit operations and supports high-fidelity two-qubit operations without spurious spectator errors. 

Two-qubit gates between fluxonium qubits are realized in multiple ways, either by direct capacitive~\cite{nesterovMicrowaveactivatedControlled$Z$Gate2018, nesterovCNOTGatesFluxonium2022,ficheuxFastLogicSlow2021, xiongArbitraryControlledphaseGate2022, baoFluxoniumAlternativeQubit2022, doganTwoFluxoniumCrossResonanceGate2023} or inductive coupling~\cite{maNativeApproachControlled$Z$2024,lin24DaysStableCNOT2025a}, and alternatively via tunable couplers with capacitive~\cite{moskalenkoHighFidelityTwoqubit2022a, simakovCouplerMicrowaveActivatedControlledPhase2023, dingHighFidelityFrequencyFlexibleTwoQubit2023a, zhanScalableFluxoniumQuantum2026a, singhFastMicrowavedrivenTwoqubit2026, zhaoScalableFluxoniumqubitArchitecture2026} or inductive couplings~\cite{weissFastHighFidelityGates2022,zhangTunableInductiveCoupler2024}.
In these architectures, CZ-gate fidelities above 99.9\,\% and residual ZZ interactions at the few-kHz level have been achieved~\cite{dingHighFidelityFrequencyFlexibleTwoQubit2023a, zhangTunableInductiveCoupler2024, lin24DaysStableCNOT2025a, zhanScalableFluxoniumQuantum2026a}.
However, extending these architectures to a connectivity-four lattice required by the surface code introduces competing demands: the coupler elements must remain spectrally isolated from both qubits and from each another, and spectator errors---the degradation of gate fidelity caused by the always-on presence of additional coupled qubits---must be systematically mitigated~\cite{krinnerBenchmarkingCoherentErrors2020, sundaresanReducingUnitarySpectator2020, zajacSpectatorErrorsTunable2021, caiImpactSpectatorsTwoQubit2021,takitaDemonstrationWeightFourParity2016, chanSystemLevelDesignScalable2026, huang2026explorationfluxoniumparameterscapacitive, zwanenburgCrosstalkMultiQubitFluxonium2026}.
Moreover, due to the small total capacitance of the fluxonium in comparison to transmons, the capacitance budget to realize strong couplings is severely constrained~\cite{guanCapacitiveLoadingTwodimensional2026, zhaoExtensibleFluxoniumArchitecture2026}.
These challenges have so far prevented a demonstration of multi-qubit fluxonium entanglement within a connectivity of four or greater and with gate fidelities compatible with fault-tolerance requirements.
A coupling strategy alternative to the previously mentioned tunable couplers employs superconducting resonators as mediating elements between fluxonium qubits, realizing a fluxonium-resonator-fluxonium (FRF) architecture~\cite{rosenfeldHighFidelityTwoQubitGates2024}. 
Resonator couplers offer several advantages over tunable elements in the context of high-connectivity lattices. 
First, resonators do not require non-linear inductances and can therefore be fabricated without Josephson junctions, which significantly improves parameter targeting and decoherence rates, a known limitation of tunable couplers~\cite{xuHighFidelityHighScalabilityTwoQubit2020, marxer999FidelitySingleQubit2025}.
Second, their flexible footprint is well-suited for the strict routing requirements of larger integrated devices, potentially enabling long-range coupling~\cite{majerCouplingSuperconductingQubits2007a,dengLongRange$ZZ$Interaction2025}. 
Third, the harmonic resonator mode can be exploited as a resource for multi-qubit gate protocols such as the resonator-induced phase (RIP) gate~\cite{pechalGeometricPhaseNonadiabatic2012, crossOptimizedPulseShapes2015,puriHighFidelityResonatorInducedPhase2016}, which produces conditional phase shifts on pairs of qubits via a common bus without requiring a tunable frequency element in the coupler circuit.
While RIP gates have been demonstrated with high fidelity in transmon architectures~\cite{paikExperimentalDemonstrationResonatorInduced2016, kumphDemonstrationRIPGates2024} and two-qubit fluxonium devices~\cite{xiongScalableLowoverheadSuperconducting2026}, its realization in multi-qubit fluxonium processors has not yet been reported.

Here, we present a fluxonium processor that employs lumped-element resonator couplers to, for the first time, realize a connectivity-four unit cell. 
We characterize the single-qubit performance of the processor, examine the suppression of residual ZZ interactions across all coupled qubit pairs and benchmarking single-qubit gate fidelities via simultaneous randomized benchmarking (RB)~\cite{emersonScalableNoiseEstimation2005, magesanScalableRobustRandomized2011} (Sec.~\ref{section:single-qubit}). 
To enable high-fidelity two-qubit operations, we employ a refocused RIP gate that mitigates spectator errors arising from the multi-qubit connectivity of the unit cell, and characterize two-qubit gate fidelities using interleaved randomized benchmarking (IRB)~\cite{magesanEfficientMeasurementQuantum2012}. 
Finally, we demonstrate multi-qubit entanglement within the unit cell by preparing Greenberger-Horne-Zeilinger (GHZ) states~\cite{greenbergerBellsTheoremInequalities1990} of up to five qubits and evaluate their fidelity via quantum state tomography. 
\section{\label{section:architecture}Device architecture}
\begin{figure}
    \includegraphics[]{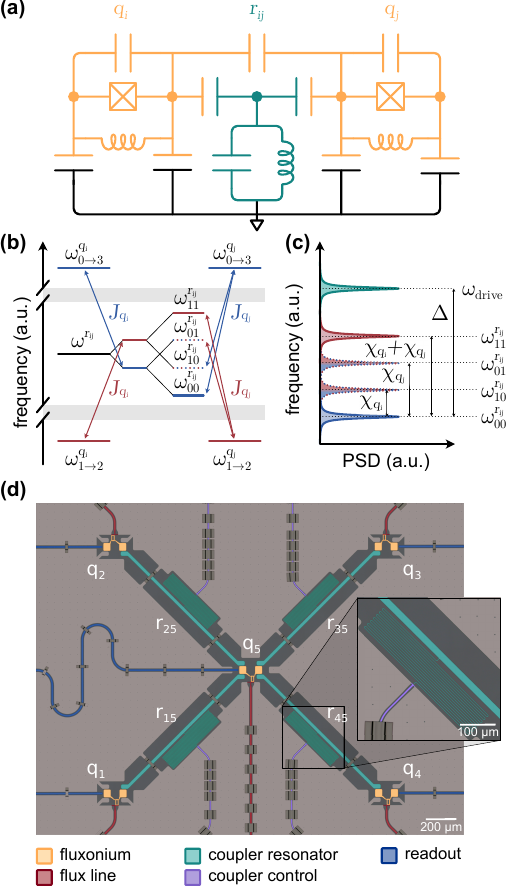}
    \caption{\textbf{Overview of the chip architecture.} (a) Simplified circuit diagram of the FRF architecture, where qubits $q_i$ and $q_j$ are in a floating configuration and the coupler $r_{ij}$ is in a grounded configuration, resulting in the suppression of the static ZZ-interaction. (b) Spectral diagram of qubit and coupler transitions leading to the dispersive shift used for the RIP gate. For qubit $q_i$ and $q_j$ respectively, the $\ket 1\rightarrow\ket 2$ and $\ket 0\rightarrow\ket 3$ transitions dispersively shift the resonator frequency $\omega^{r_{ij}}$, resulting in four spectral lines conditional on the four computational states. Blue arrows indicate coupling when a qubit is in $\ket 0$, red arrows when it is in $\ket 1$. (c) Setup of the resonator drive. The coupler is driven off-resonantly with detuning $\Delta$ to all four transition frequencies, where either $\omega_\mathrm{drive}\ll\omega^{r_{ij}}_{00}$ or $\omega_\mathrm{drive}\gg\omega^{r_{ij}}_{11}$. The power-spectral-density (PSD) and frequency spacings are not to scale. (d) False-color micrograph of the connectivity-four device. Each qubit is capacitively coupled to its next neighbor by a lumped-element resonator and to its individual readout resonator realized by a quarter-wavelength CPW resonator. Qubits and couplers are controlled via inductively coupled flux lines. }
    \label{figure1}
\end{figure}
To realize the FRF architecture outlined above, we consider a pair of neighboring fluxonium qubits, $\{q_i,q_j\}$, capacitively coupled to a shared lumped-element resonator $r_{ij}$ that mediates their interaction. 
We model this three-mode system following Rosenfeld \textit{et al.}~\cite{rosenfeldHighFidelityTwoQubitGates2024} using the Hamiltonian $\hat{H} = \hat{H}_0 + \hat{H}_\mathrm{int}$.
The uncoupled Hamiltonian is given by
\begin{equation}
\begin{split}
    \hat{H}_0 = &\sum_{i} \left[ 4E_{\mathrm{C},q_i}\hat{n}_{q_i}^2 + \frac{E_{\mathrm{L},q_i}}{2}\hat{\phi}_{q_i}^2 + E_{\mathrm{J},q_i}\cos\hat{\phi}_{q_i} \right]\\ 
    + &\sum_{\{i,j\}}\hbar\omega^{r_{ij}}\hat{a}^\dagger_{r_{ij}}\hat{a}_{r_{ij}},
\end{split}
\end{equation}
and the interaction Hamiltonian reads
\begin{equation}
    \hat H_\mathrm{int} = 
    \sum_{\{i,j\}} J_{q_iq_j}\,\hat n_{q_i}\hat n_{q_j}
    +J_{q_ir_{ij}}\,\hat n_{q_i}\hat n_{r_{ij}}
    +J_{q_jr_{ij}}\,\hat n_{q_j}\hat n_{r_{ij}}.
\end{equation}
Here, $\hat{n}_{q_i}$ and $\hat{\phi}_{q_i}$ are the charge and phase operators of fluxonium $q_i$, with circuit parameters $E_{\mathrm{C},{q_i}}$, $E_{\mathrm{L},{q_i}}$, and $E_{\mathrm{J},{q_i}}$ denoting the charging, inductive, and Josephson energies, respectively. 
All qubits are flux-biased at the lower sweet spot where the external flux is one half flux quantum, $\phi_\mathrm{ext} = 0.5\,\Phi_0$. The coupler resonators are specified by their resonance frequency $\omega^\mathrm{r_{ij}}$ and photon creation and annihilation operators $\hat{a}^\dagger_{r_{ij}}$ and $\hat{a}_{r_{ij}}$. 
The interaction Hamiltonian captures the pairwise capacitive couplings between all three modes with the coupling strength $J_{kl}$ between the circuit elements $k$ and $l$. A simplified circuit diagram of a two-qubit subset is depicted in Fig.~\ref{figure1}\,(a).

All fluxonium qubits are designed to the target parameters $E_\mathrm{C}/h = 1.5\,\mathrm{GHz}$, $E_\mathrm{J}/h = 6.5\,\mathrm{GHz}$, and $E_\mathrm{L}/h = 0.9\,\mathrm{GHz}$, yielding a qubit transition frequency of $\omega_\mathrm{q}/2\pi \approx 350\,\mathrm{MHz}$.
The physical device parameters are summarized in Table~\ref{tab:single-qubit-parameters}.
To execute two-qubit gates via the coupler resonator, we exploit the coupling to transitions out of the computational subspace to generate a state-dependent dispersive shift on the resonator frequency. 
Specifically, we design $\omega^{r_{ij}}$ to lie above the $|1\rangle \to |2\rangle$ transition and below the $|0\rangle \to |3\rangle$ transition of each qubit, as outlined in Fig.~\ref{figure1}\,(b). 
In this configuration, the resonator frequency is shifted downward when $q_i$ occupies $|0\rangle$, through coupling to the $|0\rangle \to |3\rangle$ transition, and upward when $q_i$ occupies $|1\rangle$, through coupling to the $|1\rangle \to |2\rangle$ transition. 
The net result is a positive dispersive shift $\chi_{q_i}$ per qubit. To maintain the coupler in the dispersive regime, we require $J_{q_ir_{ij}}|\langle k | \hat{n}_{q_i} | l \rangle| \ll |\omega^{r_{ij}} - \omega^{q_i}_{kl}|$ for all relevant transitions~\cite{dingNovelGatesSuperconducting2023}.
Taking into account both $q_i$ and $q_j$, the resonator acquires four distinct frequencies depending on the computational state of the qubit pair:
\begin{equation}
  \begin{split}
    \omega^{r_{ij}}_{00}&,\quad 
    \omega^{r_{ij}}_{10} = \omega^{r_{ij}}_{00} + \chi_{q_i},\\
    \omega^{r_{ij}}_{01} = \omega^{r_{ij}}_{00} + \chi_\mathrm{q_j}&,\quad
    \omega^{r_{ij}}_{11} = \omega^{r_{ij}}_{00} + \chi_{q_i} + \chi_{q_j}.
\end{split}  
\end{equation}
This virtual interaction scheme is possible due to the selection rules the fluxonium circuit exhibits at the lower sweet-spot, where even-numbered transitions like $\ket 0\rightarrow\ket 2$ and $\ket 1\rightarrow\ket 3$ are forbidden.

A controlled-Z (CZ) gate is then implemented by driving the fundamental mode of the coupler resonator with a shaped microwave drive~\cite{pechalGeometricPhaseNonadiabatic2012, crossOptimizedPulseShapes2015}, as sketched in Fig.~\ref{figure1}(c).
This leads to a qubit state dependent excursion of the resonator in its phase space and a subsequent state-specific phase accumulation.
The drive is applied at a frequency detuned by $\Delta = \omega_\mathrm{drive} - \omega^{r_{ij}}_{00}$ from the resonator frequency, where $\Delta$ is chosen such that the drive is either below $\omega^{r_{ij}}_{00}$ or above $\omega^{r_{ij}}_{11}$, avoiding placement between any pair of the four state-dependent resonator frequencies.
This detuning is required to return the resonator to the vacuum state at the end of the gate~\cite{crossOptimizedPulseShapes2015}.
The physical implementation of the unit cell is shown in Fig.~\ref{figure1}(d). 
Each fluxonium qubit is capacitively coupled to a coplanar waveguide (CPW) readout resonator and is controlled through an inductively coupled flux line used for both flux biasing and single-qubit drives. 
The coupler resonators are realized as grounded lumped-element LC resonators, in which a 2\,\textmu m niobium wire serves as the inductive element. 
This design yields a characteristic impedance of approximately 150\,$\Omega$ required to achieve a sufficiently small capacitive loading of each fluxonium qubit to reach a connectivity of four. 
The high resonator impedance simultaneously enables large qubit-resonator coupling strengths and therefore large dispersive shifts $\chi_{q_i}$ (cf. Table~\ref{tab:coupler-parameters}), while the grounded resonator geometry suppresses the static ZZ interaction $\xi_\mathrm{ZZ}$ between the fluxonium qubits to below 1\,kHz~\cite{rosenfeldHighFidelityTwoQubitGates2024}.
Although higher impedances are beneficial for relaxing the capacitance budget further, it limits the available capacitance of the coupler itself, in turn constraining the wire routing of the processor by limiting the couplers physical size.
The drive line of each coupler is coupled inductively to the grounded end of the niobium wire.
This arrangement provides a coupling to the resonator mode while coupling only weakly to other circuit elements.
The resulting parameters of the five-qubit unit cell, including the qubit frequencies $\omega_\mathrm q$, resonator frequencies $\omega^\mathrm{r_{ij}}_{00}$ and the dispersive shifts $\chi_{q_i}$ extracted from state-dependent resonator spectroscopy, are summarized in Table~\ref{tab:single-qubit-parameters} and Table~\ref{tab:coupler-parameters}.
\begin{figure}[t]
    \includegraphics[]{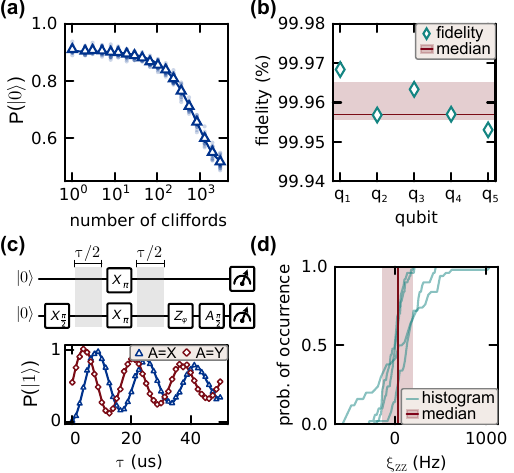}
    \caption{\textbf{Qubit performance.} (a) Exemplary single-qubit RB of $q_1$ with up to 3000 Cliffords. (b) Average single-qubit gate fidelity measured simultaneously on all qubits. The error is smaller than the marker size (cf. Table \ref{tab:single-qubit-parameters}). (c) Refocusing sequence for measuring static $\xi_\mathrm{ZZ}$, where a single qubit is initialized in a superposition state, followed by a wait time $\tau/2$, a refocusing pulse on both qubits. For a robust fit, we apply a virtual frequency through a virtual-$Z$ rotation before the second $\pi/2$-pulse and extract the conditional phase from the deviation to the virtual oscillation. (d) Histogram of 50 repetitions of the procedure shown in (c) for all connected qubit pairs, resulting in a median $\xi_\mathrm{ZZ}=30(215)$\,Hz. }
    \label{figure2}
\end{figure}
\section{\label{section:single-qubit}Qubit performance}
To assess the performance of each qubit, we record their relaxation time $T_1$, Ramsey- and spin-echo decoherence times $T_2^*$ and $T_2^\mathrm e$, as well as single-qubit gate fidelities $\mathcal F_\mathrm{1q}$.
The measured qubit parameters are summarized in Table~\ref{tab:single-qubit-parameters} and the measurement setup is detailed in Appendix~\ref{appendix:setup}.
We implement single-qubit gates with 32\,ns Gaussian-shaped pulses followed by a 16\,ns delay, which accounts for signal reflections in the flux line caused by the diplexing of the bias- and drive-lines. 
We measure the average fidelity of 48\,ns single-qubit gates using RB, exemplified by Fig.~\ref{figure2}\,(a), on all qubits simultaneously. Using the gate set $\mathcal{G}=\{I, X_{\pm\pi}, Y_{\pm\pi}, X_{\pm\pi/2}, Y_{\pm\pi/2}\}$ resulting in 1.875 average gates per Clifford, we extract $\mathcal{F_\mathrm{1Q}}$ [see Fig.~\ref{figure2}\,(b)] from 30 different random Clifford sequences.
The resulting median fidelity of 99.957(5)\,\% approaches the coherence limit of 99.96(1)\,\% of the device determined by the coherence times stated in Table~\ref{tab:single-qubit-parameters} and the limit discussed in Refs.~\cite{dingHighFidelityFrequencyFlexibleTwoQubit2023a, pedersenFidelityQuantumOperations2007}.
These results demonstrate that the FRF architecture supports parallel single qubit operations at high fidelity close to the coherence limit of the device.

To characterize the idling performance of the unit cell, we measure $\xi_\mathrm{ZZ}$ between coupled qubit pairs.
We use a joint amplification of ZZ interaction (JAZZ) sequence~\cite{liRealizationHighFidelityCZ2024} depicted in Fig.~\ref{figure2}\,(c). 
The sequence prepares one qubit in $\ket 0$ and the other in an equal superposition state, followed by a delay time $\tau/2$, after which the qubit states are inverted by an $X_\pi$-pulse applied to both qubits, followed by another delay time $\tau/2$ and a final measurement.
Due to the echo pulse on both qubits, the single-qubit phase accumulated during the first and second delay time cancel, leaving only the conditional phase due to ZZ-crosstalk.
To make the measurement robust to fit uncertainties, we repeat the sequence with $X_{\pi/2}$- and $Y_{\pi/2}$-pulses as the final pulse, denoted by the index $i\in\{0,1\}$.
Furthermore, as the ZZ interaction is weak, we apply a virtual phase rotation before the last $\pi/2$-pulse as a function of the delay time $\tau$ resulting in a virtual oscillation with frequency $f_\mathrm{V}=n/T$, where $T$ is the maximum sequence delay and $n=3$ is chosen such that enough oscillations are visible in the dataset. 
$\xi_\mathrm{ZZ}$ is then determined as twice the difference between the observed frequency to the virtual frequency $\xi_\mathrm{ZZ}=2(f-f_\mathrm{V})$.
As the observed $\xi_\mathrm{ZZ}$ in our device is much smaller than the qubit decoherence rate $\Gamma_2\sim10$\,kHz, we record 50 JAZZ sequences for each qubit pair and determine the rate from the integrated histograms in Fig.~\ref{figure2}\,(d), resulting in a median $\xi_\mathrm{ZZ}=30(215)$\,Hz, reaffirming the sufficient suppression of residual ZZ interactions in the FRF architecture~\cite{xiongScalableLowoverheadSuperconducting2026} for surface code operation~\cite{zhouSurfaceCodeError2025}.
\section{\label{section:two-qubit}CZ-gate performance and spectator errors}
\begin{figure}
    \includegraphics[]{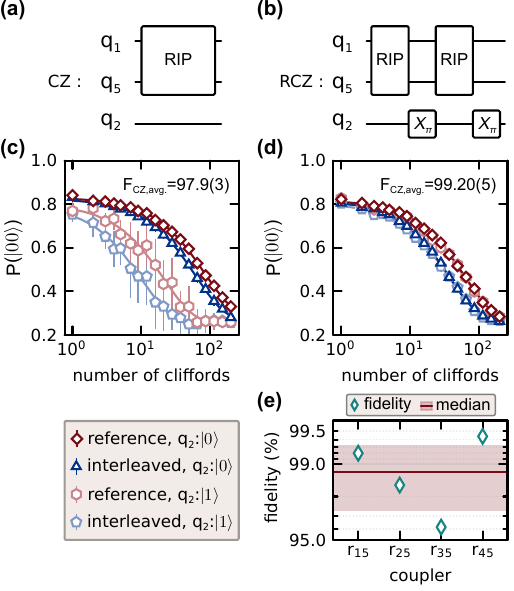}
    \caption{\textbf{Spectator error mitigation.} (a) Configuration of a regular RIP gate between two qubits $q_1$ and $q_5$ with no operation on the spectator qubit $q_2$ results in a spectator state-dependent error when $q_2$ is in $\ket 1$, as characterized by IRB (c). The gate is calibrated for the case where $q_2$ is in $\ket 0$. (b) Refocusing RIP gate. The RIP pulse is split into two sections, split by a $\pi$-pulse on $q_2$, resulting in a cancellation of conditional phases caused by the spectator qubit and no dependence of the CZ-gate fidelity in IRB (d). (e) Average two-qubit gate fidelity using the refocusing sequence on different qubit pairs.} 
    \label{figure3}
\end{figure}
To realize RIP-based CZ gates, we apply a detuned 128\,ns cosine-shaped pulse to the coupler which is optimized by a tune-up procedure described in Appendix~\ref{appendix:tune-up}.
Following the tune-up, we determine the CZ-gate fidelity by IRB, shown in Fig.~\ref{figure3}\,(c) exemplarily for coupler $r_{15}$ (dark colors) with a fidelity of $\mathcal F_\mathrm{CZ}=99.68(5)$\,\%.
The gate performance is however not independent of the state of neighboring qubits.
If a neighboring qubit, here chosen to be $q_2$, is initialized in $\ket 1$ as opposed to $\ket 0$ prior to the IRB sequence (Fig.~\ref{figure3}\,(c), bright colors), the arising spectator errors reduce the fidelity to $\mathcal F_\mathrm{CZ}=96.1(6)$\,\%, with an average fidelity of 97.9(3)\,\% across both IRB sequences.
To cancel this state dependence, we utilize a Hamiltonian refocusing protocol~\cite{paikExperimentalDemonstrationResonatorInduced2016,takitaDemonstrationWeightFourParity2016, glaserControlledControlledPhaseGatesSuperconducting2023} shown in Fig.~\ref{figure3}\,(b), where the RIP interaction is split into two pulses.
After each pulse, the spectator qubit states are inverted by a $X_\pi$-pulse which causes their contribution to the conditional phase to invert, and thus refocusing the conditional phase accumulation to zero.
At the same time, the refocusing sequence results in an identity operation on the spectator qubits.
We implement this protocol using two 96\,ns pulses on the coupler using the same tune-up procedure and measure again a spectator state dependent IRB. 
As Fig.~\ref{figure3}\,(b) demonstrates, the refocused gate removes the spectator state dependence up to statistical uncertainty and increases the average fidelity across both IRB sequences to $99.20(5)$\,\%.
Moreover, the success of the echo sequence provides strong evidence that the spectator error is of a ZZ-type.
We employ this sequence for all qubits and apply the refocusing pulse to all next-neighbor qubits, resulting in a median CZ-gate fidelity of 98.8(12)\,\% as detailed in Fig.~\ref{figure3}\,(e) and Table~\ref{tab:coupler-parameters}.
This median value is dominated by the relatively low fidelity of 96.2(2)\,\% on coupler $r_{35}$ which we attribute to an increased photon loss during the gate operation.
Due to the couplers small dispersive shift of $\chi=0.81$\,MHz, the detuning of the drive tone must be decreased to execute the RIP gate within the specified pulse time, in turn increasing decoherence due to the finite linewidth of the coupling resonator~\cite{crossOptimizedPulseShapes2015, xiongScalableLowoverheadSuperconducting2026}.
\begin{table}
    \centering
    \begin{tabular}{ccccccc}
      qubit   &  $\omega_\mathrm q$ (MHz) & $T_1$ (\textmu s) & $T_2^*$ (\textmu s) & $T_2^\mathrm e$ (\textmu s) & $\mathcal F_\text{1Q}$ (\%)\\
         \hline
         \hline
         $q_1$&433&87(38)&53(2)&68(4)&99.968(1) \\
         $q_2$&441&65(12)&52(4)&87(12)&99.957(1) \\
         $q_3$&493&57(41)&70(40)&82(4)&99.963(1) \\
         $q_4$&399&73(34)&43(5)&69(4)&99.957(1) \\
         $q_5$&462&62(11)&48(27)&65(4)&99.953(1) \\
         \hline
    \end{tabular}
    \caption{\textbf{Collection of qubit parameters.} $T_1$, $T_2^*$ and $T_2^\mathrm e$ represent median values and standard deviations of 50 decay, ramsey and echo experiments, respectively. 
    Due to the low frequency of fluxonium qubits, $T_1$ includes contributions from spontaneous relaxation $T_1^\downarrow$ and thermally induced excitation $T_1^\uparrow$ resulting in the total relaxation time $1/T_1=1/T_1^\uparrow+1/T_1^\downarrow$.
    We record both cases by alternating the state preparation between $\ket 0$ and $\ket 1$ and measuring the decay.}
    \label{tab:single-qubit-parameters}
\end{table}
\begin{table}[]
    \centering
    \begin{tabular}{ccccccc}
      coupler   &  $\omega^{r_{ij}}_{00}$ (GHz) & $\chi_{q_i}$ (MHz) & $\chi_{q_j}$ (MHz) & $\xi_\mathrm{ZZ}$ (Hz) & $\mathcal F_\text{CZ}$ (\%)\\
         \hline
         \hline
         $r_{15}$&7.358&19.75&11.55&60(355)&99.20(5) \\
         $r_{25}$&7.494&20.3&53.9&-6(111)&98.4(1) \\
         $r_{35}$&7.635&-19.2&0.81&2(95)&96.2(2) \\
         $r_{45}$&7.793&25.4&40.6&131(145)&99.44(3) \\
         \hline
    \end{tabular}
    \caption{\textbf{Collection of coupler parameters.} Resonator frequencies are determined using state-dependent resonator spectroscopy, median $\xi_\mathrm{ZZ}$ and standard deviations are determined using 50 JAZZ sequences and $\mathcal F_\mathrm{CZ}$ are determined by 30 IRB sequences with a maximum length of 200 Cliffords. The dispersive shift is ordered such that $\chi_{q_i}$ is the shift of the first coupler index and $\chi_{q_j}$ the shift of the second coupler index.}
    \label{tab:coupler-parameters}
\end{table}
\section{\label{section:ghz-states}Greenberger-Horne-Zeilinger (GHZ) states}
\begin{figure}
    \includegraphics[]{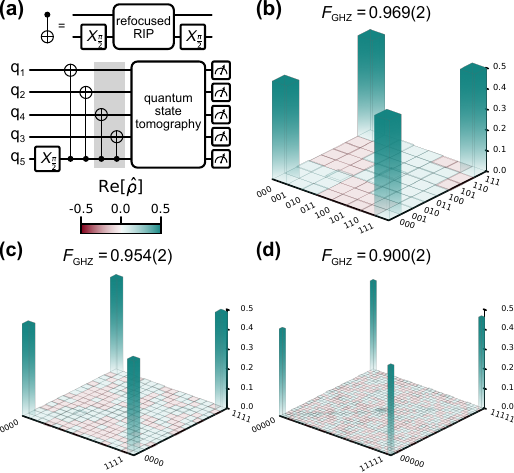}
    \caption{\textbf{GHZ states}. (a) Circuit diagram to prepare and tomographically determine the fidelity of a three-qubit(b) four-qubit (c) and five-qubit (d) GHZ-state. $q_5$ is prepared in an equal superposition, followed by effective CNOT gates to all other qubits. This is followed by a state tomography circuit which prepares all Pauli-correlators with subsequent measurement. CNOT gates marked in grey are only applied for the four- and five-qubit GHZ state. Each $\hat\rho$ is recorded with 30720 single shots. All fidelities shown are corrected for SPAM errors by classical detector error correction.}
    \label{figure4}
\end{figure}
With single- and two-qubit gates characterized at high fidelity, we now use the resulting gate set to generate and verify genuine multi-qubit entanglement across the connectivity-four unit cell.
For this purpose, GHZ states are a natural choice as they provide a stringent benchmark for both single- and two-qubit operations~\cite{zhanScalableFluxoniumQuantum2026a}.
In fact, genuine multipartite entanglement is verified by measuring a state fidelity exceeding 50\,\%~\cite{guhneSeparabilityCriteriaGenuine2010}.
We prepare and benchmark three-, four- and five-qubit GHZ states using quantum state tomography~\cite{jamesMeasurementQubits2001}.
As shown in Fig.~\ref{figure4}\,(a), we first prepare an equal superposition state on $q_5$, followed by CNOT gates to all outer qubits, composed from CZ gates and single-qubit rotations.
To achieve the highest state preparation fidelities, we choose the smaller GHZ states [$(q_1, q_2,q_5)$ and $(q_1, q_2, q_4,q_5)$ for the 3 and 4 qubit GHZ states, respectively] to contain qubit pairs with the highest CZ-gate fidelities.
Following the state preparation, we measure all combinations of $N$-qubit Pauli correlators $\langle\hat P^{\otimes N}\rangle$ with $\hat P\in\{\hat X, \hat Y, \hat Z\}$ and $N\in\{3,4,5\}$ by applying the corresponding single-qubit rotations followed by simultaneous single-shot readout of all qubits.
Before reconstructing the density matrix, we apply a classical measurement-error correction to the recorded data~\cite{maciejewskiMitigationReadoutNoise2020, aasenReadoutErrorMitigated2024}.
The detector error is characterized by the matrix $M$ defined via $\mathbf p_\mathrm{exp}=M\mathbf p_\mathrm{ideal}$, which maps the ideal measurement probabilities $\mathbf p_\mathrm{ideal}$ onto the experimentally observed probabilities $\mathbf p_\mathrm{exp}$.
We determine $M$ by preparing and measuring all computational states and invert it to recover the ideal probabilities, $\mathbf p_\mathrm{ideal}=M^{-1}\mathbf p_\mathrm{exp}$.
From the measurement-corrected probabilities, we reconstruct a physical density matrix using maximum-likelihood estimation~\cite{hradilQuantumstateEstimation1997, banaszekMaximumlikelihoodEstimationDensity1999}, yielding tomographic fidelities of 96.9(2)\,\%, 95.4(2)\,\% and 90.0(2)\,\% for the three-, four-, and five-qubit GHZ states shown in Figs.~\ref{figure4}\,(b)--(d) respectively.
For all measured states, the state fidelity before measurement error correction is above 60\,\% as discussed in Appendix~\ref{appendix:qst}.
Consistent with the two-qubit characterization in Sec.~\ref{section:two-qubit}, the reconstructed GHZ density matrices show no signatures of spectator-induced coherent errors such as pairwise off-diagonal elements of opposing sign or with significant imaginary components, confirming that the refocused RIP gate suppresses spectator errors effectively across multi-qubit circuits.
\section{\label{section:discussion}Conclusion \& Outlook}
In summary, we have realized a five-qubit fluxonium processor based on a FRF architecture that realizes a connectivity-four unit cell compatible with a surface-code lattice. 
The use of high-impedance, grounded lumped-element resonator couplers suppresses the static ZZ interaction between coupled fluxonium pairs to a median rate of $30(215)$\,Hz, while supporting parallel single-qubit gates at a median fidelity of 99.957(5)\,\% in simultaneous randomized benchmarking. 
Spectator errors arising from the multi-qubit connectivity of the unit cell are mitigated by a refocused RIP gate, which restores the two-qubit gate fidelity to a median of 98.8(12)\,\% in interleaved randomized benchmarking and removes the dependence on the spectator-qubit state to within statistical uncertainty. 
With the resulting gate set, we prepare GHZ states of three, four and five qubits with tomographic fidelities of 96.9(2)\,\%, 95.4(2)\,\% and 90.0(2)\,\% respectively, demonstrating, to our knowledge, the first multi-qubit entangled states in fluxonium qubits with a connectivity of four~\cite{mazhorinNativeCCZGate2026}, as well as the highest GHZ fidelities using fluxonium qubits per respective qubit count~\cite{zhanScalableFluxoniumQuantum2026a}.

The device is currently limited by the variation in two-qubit gate fidelity across the four coupled pairs, with the pair $q_3q_5$ reaching only 96.2(2)\,\% in IRB. 
This variation is consistent with the corresponding spread in dispersive shifts and coupler frequencies (cf. Table~\ref{tab:coupler-parameters}), which set the achievable gate speed and the leakage budget of the RIP interaction. 
While this limit is present in our specific device, it is not a limitation of the FRF architecture itself, as the remaining qubit pairs do not show such a limitation.
Improving the uniformity of these parameters is therefore a primary target for future fabrication and design iterations of the FRF architecture, and will allow the intrinsic gate-fidelity ceiling of the RIP interaction to be mapped out. 
A second limitation concerns the RIP gate duration of 128\,ns (or $2\times96$\,ns for the refocused RIP gate), which is comparatively slow relative to the fastest two-qubit gates demonstrated in fluxonium architectures based on direct or tunable couplings. 
Reducing this duration through pulse shaping~\cite{crossOptimizedPulseShapes2015,xiongScalableLowoverheadSuperconducting2026} without reintroducing leakage or spectator errors is an important step toward bringing the FRF architecture in line with the gate speeds required for scalable, surface-code-compatible fluxonium processors.

We speculate that the observed spectator errors are caused by residual coupling between neighboring couplers as described in Appendix~\ref{appendix:design-parameters}, which in combination with their relatively small mutual detuning mediate a state-dependent shift. 
The refocused RIP gate suppresses such errors without requiring modifications of the quantum processor, demonstrating that multi-qubit connectivity in the FRF architecture is compatible with high-fidelity two-qubit operations. 
However, the refocusing sequence requires the spectator qubit to remain idle and addressable for the $X_\pi$-pulse, which prevents parallel execution of two-qubit gates on neighboring qubit pairs. 
In the context of the surface code, this translates into stabilizer measurements distributed across two cycles rather than one.
We are confident that increasing the coupler-coupler detuning and reducing the residual coupling through careful optimization of the inter-coupler capacitances should mitigate spectator errors at the hardware level and thereby recover full parallelism in future device generations.

Finally, the present device is limited by its readout performance, with a median single-shot assignment fidelity of $94_{-3}^{+1}$\,\%. 
The classical detector error correction recovers the state fidelities as reported in Sec.~\ref{section:ghz-states}, but an assignment error at this level precludes high-fidelity syndrome extraction required for quantum error correction. 
Since the readout circuitry of this device was not optimized alongside the coupler design, this is not a fundamental limitation of fluxonium-based architectures~\cite{nesterovMeasurementinducedStateTransitions2024, chappleMeasurementinducedStateTransitions2026, botharaHighfidelityQNDReadout2025a, bistaReadoutinducedLeakageFluxonium2026, singhImpactJosephsonJunctionArray2025}, and established techniques such as Purcell-filtered readout resonators~\cite{reedFastResetSuppressing2010, jeffreyFastAccurateState2014}, an optimized combination of dispersive shift and resonator linewidth~\cite{walterRapidHighFidelitySingleShot2017, krantzQuantumEngineersGuide2019} or dynamic readout protocols~\cite{stefanskiImprovedFluxoniumReadout2024} will be integrated into future device generations.

The FRF architecture exhibits several intrinsic strengths as a building block for scalable fluxonium processors. 
The grounded high-impedance resonator coupler provides robust suppression of the residual ZZ interaction across all coupled pairs and is straightforward in design: as a passive linear element, the coupler is easily targetable in frequency, shifting the precision requirement onto the placement of the higher excited states of the fluxonium that mediate the dispersive interaction. 
The CZ gate is implemented as a purely microwave-driven operation on the coupler, which removes the need to calibrate flux-pulse cross-talk between neighboring qubits---a calibration step that typically introduces substantial overhead in flux-activated two-qubit gate schemes. 
In addition, the present work shows that the spectator errors arising from the multi-qubit connectivity can be addressed by a straightforward refocusing protocol at the pulse level. Taken together, the robust ZZ suppression, the calibration-light microwave control, and the demonstrated mitigation of spectator errors establish the FRF unit cell as a viable building block for densely connected fluxonium processors. 
Beyond the surface code, the natural extensibility of resonator-mediated couplings make the FRF architecture a promising platform for codes requiring enhanced qubit connectivity~\cite{xiongScalableLowoverheadSuperconducting2026} including qLDPC codes~\cite{breuckmannQuantumLowDensityParityCheck2021, panteleev2022asymptoticallygoodquantumlocally, bravyiHighthresholdLowoverheadFaulttolerant2024}, and thereby provides a scalable path toward low-overhead fluxonium-based quantum error correction.
\section*{Acknowledgements}
This work received financial support from  the German Federal Ministry of Education and Research via the funding program ’Quantum technologies - from basic research to the market’ under contract number 13N15680 (GeQCoS) and under contract number 13N16188 (MUNIQC-SC), by the Deutsche Forschungsgemeinschaft (DFG, German Research Foundation) via the Germany’s Excellence Strategy EXC-2111-390814868 ‘MCQST’ as well as by the European Union by the EU Flagship on Quantum Technology HORIZON-CL4-2022-QUANTUM-01-SGA project 101113946 OpenSuperQPlus100. The research is part of the Munich Quantum Valley, which is supported by the Bavarian state government with funds from the Hightech Agenda Bayern Plus.

\appendix
\section{\label{appendix:design-parameters}Design Parameters}
The fluxonium qubits are designed with $E_\mathrm{C}/h = 1.5\,\mathrm{GHz}$, $E_\mathrm{J}/h = 6.5\,\mathrm{GHz}$ and $E_\mathrm{L}/h = 0.9\,\mathrm{GHz}$ where $h$ denotes Planck's constant, yielding a qubit transition frequency $\omega_\mathrm{q}/2\pi \approx 350\,\mathrm{MHz}$ and plasmon transition frequencies $\omega_{12}/2\pi = 5.81\,\mathrm{GHz}$ and $\omega_{03}/2\pi = 8.90\,\mathrm{GHz}$.
The relatively large splitting between the two plasmon transitions accommodates four coupler resonators and five readout resonators between them with sufficient frequency spacing.
In particular, the readout resonators are located between 6.35\,GHz to 6.55\,GHz, at least 1\, GHz detuned from the nearest coupler resonator to prevent spurious interactions.
To reach dispersive shifts above $20\,\mathrm{MHz}$ while remaining in the dispersive regime, we detune each coupler by at least $1\,\mathrm{GHz}$ from the nearest plasmon transition.
The corresponding coupler design parameters are summarized in Table~\ref{tab:coupler-design-parameters}.
Each coupler is targeted to a characteristic impedance $Z$ above $150\,\Omega$, which is required to reach sufficiently large qubit-resonator couplings while remaining within the capacitance budget imposed by the four couplers attached to the central qubit~$q_5$.
The capacitive charging energy is identical across all four couplers, and the mode frequency $\omega^{r_{ij}}/2\pi$ is set by the length of the inductive shunt.
Compared to the design frequencies given in Table~\ref{tab:coupler-design-parameters}, the measured device frequencies summarized in Table~\ref{tab:coupler-parameters} show a systematic shift of $\sim200$\,MHz, which we attribute to a systematic decrease of the couplers inductor width during ground plane fabrication.
\begin{table}[]
    \centering
    \begin{tabular}{ccccc}
         coupler & $E_\mathrm{C}/h$ (GHz) & $E_\mathrm{L}/h$ (GHz) & $Z$ ($\Omega$) & $\omega^{r_{ij}}/2\pi$ (GHz)\\
         \hline
         \hline
         $r_{15}$ & 0.149 & 48.0 & 162 & 7.569 \\
         $r_{25}$ & 0.149 & 49.9 & 158 & 7.714 \\
         $r_{35}$ & 0.149 & 51.9 & 155 & 7.854 \\
         $r_{45}$ & 0.149 & 54.1 & 152 & 7.996 \\
         \hline
    \end{tabular}
    \caption{\textbf{Coupler design parameters.} The charging energy is identical across all four couplers; the resonator mode frequency $\omega^{r_{ij}}/2\pi$ is set by the length of the inductive shunt.}
    \label{tab:coupler-design-parameters}
\end{table}
The coupling between modes is mediated by shared capacitances, resulting in the interaction Hamiltonian of the full unit cell
\begin{equation}
    \hat H_\mathrm{int} = \sum_{\{i,j\}} J_{ij}\,\hat n_i\hat n_j,
    \label{eq:coupling-full}
\end{equation}
where the element pair $\{i,j\}$ runs over all unique qubit and coupler combinations.
The coupling strengths $J_{ij}$ are obtained from electrostatic simulations using \textit{Ansys Maxwell}. 
The full coupling matrix of the five-qubit unit cell is given in Table~\ref{tab:couplings}.
The design targets a large nearest-neighbor qubit-coupler coupling of approximately $220\,\mathrm{MHz}$, together with a direct next-neighbor qubit-qubit coupling of $35$--$43\,\mathrm{MHz}$, which compensates the resonator-mediated ZZ interaction and is responsible for the residual ZZ suppression characteristic of the FRF architecture~\cite{rosenfeldHighFidelityTwoQubitGates2024}.
The provided coupling strengths are achieved while the coupling capacitance expenditure per coupler remains below 17\,\% of each qubits total available capacitance.
This direct qubit-qubit coupling arises intrinsically from the unit-cell geometry across a wide range of coupler parameters and does not require a separate qubit-qubit capacitance to be engineered.
The coupling between next-next-neighbor qubits is strongly suppressed to $\approx 1\,\mathrm{MHz}$ in this configuration, and the spurious coupling between qubits and non-neighbor couplers remains below $5\,\mathrm{MHz}$.
A residual coupling of up to $26\,\mathrm{MHz}$ persists between neighboring couplers, which we identify as the most likely cause of the spectator errors discussed in Sec.~\ref{section:discussion}~\cite{zwanenburgCrosstalkMultiQubitFluxonium2026, langeCrosstalkSuperconductingQubit2025}.

\begin{table}[]
\centering
\begin{tabular}{cccccccccc}
 & $q_1$ & $q_2$ & $q_3$ & $q_4$ & $q_5$ & $r_{15}$ & $r_{25}$ & $r_{35}$ & $r_{45}$ \\
\hline
\hline
$q_1$    & --  & 1   & 0   & 0   & 40  & 222 & 4   & 2   & 2   \\
$q_2$    & 1   & --  & 0   & 0   & 35  & 4   & 211 & 2   & 1   \\
$q_3$    & 0   & 0   & --  & 1   & 41  & 2   & 2   & 221 & 5   \\
$q_4$    & 0   & 0   & 1   & --  & 43  & 2   & 1   & 5   & 210 \\
$q_5$    & 40  & 35  & 41  & 43  & --  & 221 & 198 & 223 & 244 \\
$r_{15}$ & 222 & 4   & 2   & 2   & 221 & --  & 24  & 8   & 12  \\
$r_{25}$ & 4   & 211 & 2   & 1   & 198 & 24  & --  & 11  & 8   \\
$r_{35}$ & 2   & 2   & 221 & 5   & 223 & 8   & 11  & --  & 26  \\
$r_{45}$ & 2   & 1   & 5   & 210 & 244 & 12  & 8   & 26  & --  \\
\hline
\end{tabular}
\caption{\textbf{Full coupling matrix of the five-qubit unit cell.} All values are coupling strengths $J_{ij}/h$ in MHz, obtained from \textit{Ansys Maxwell} electrostatic simulations. The matrix is symmetric by reciprocity, $J_{ij}=J_{ji}$; both triangles are shown for ease of row- and column-wise inspection.}
\label{tab:couplings}
\end{table}

\section{\label{appendix:fabrication}Device Fabrication}
The device is fabricated following a multi-step process that combines wafer-scale subtractive niobium patterning and air bridge fabrication with subsequent die-level processing of the Josephson junctions. The fabrication sequence is designed such that all high-temperature and aggressive cleaning steps precede the deposition of the aluminum-based junctions.

We begin by preparing a $100\,\mathrm{mm}$ high-resistivity ($>10\,\mathrm{k\Omega\,cm}$) wafer in a buffered oxide etch (BOE) solution to remove lossy native silicon oxides, followed by an immediate transfer into an ultra-high vacuum (UHV) sputter deposition system (\textit{PLASSYS} MEB550 S4-I). A $150\,\mathrm{nm}$ thin film of niobium is sputter-deposited onto the wafer to form the groundplane. All macroscopic structures ($\geq 2\,\text{\textmu m}$), including the coplanar waveguide feedlines, readout resonators, flux lines, qubit pads, and the lumped-element resonator couplers, are defined using maskless optical lithography and the pattern is transferred into the niobium film via an $\mathrm{SF}_6$-based reactive ion etching process.

To suppress parasitic slotline modes and to enable the routing density required by the connectivity-four unit cell, we incorporate niobium air bridges across the feedlines and flux-control lines using the wafer-scale subtractive process described in Ref.~\cite{bruckmoserNiobiumAirBridges2026}. The wafer is then diced into individual dies for the remaining processing steps.

The Josephson junctions of the fluxonium qubits are fabricated using a Dolan-Niemeyer bridge technique~\cite{dolanOffsetMasksLiftoff1977}. We spin-coat a bilayer electron-beam resist stack and expose the junction pattern with a $200\,\mathrm{kV}$ electron-beam lithography system. After development, the junctions are formed by double-angle shadow evaporation of aluminum in a UHV chamber: a first layer of $30\,\mathrm{nm}$ of aluminum is evaporated at an angle of $+20^{\circ}$, followed by a dynamic oxidation step that defines the tunnel barrier of the Josephson contact, and a second layer of $70\,\mathrm{nm}$ of aluminum evaporated at the opposite angle of $-20^{\circ}$. The resist and excess aluminum are subsequently lifted off in N-methyl-2-pyrrolidone (NMP).

Finally, the aluminum junctions are galvanically connected to the niobium groundplane via aluminum bandages~\cite{dunsworthCharacterizationControlMitigation2017}. These are defined in a second electron-beam lithography step, followed by in-situ argon ion milling to remove native oxides and a subsequent aluminum evaporation.

\section{\label{appendix:setup}Experimental Setup}
\begin{figure}
    \centering
    \includegraphics[]{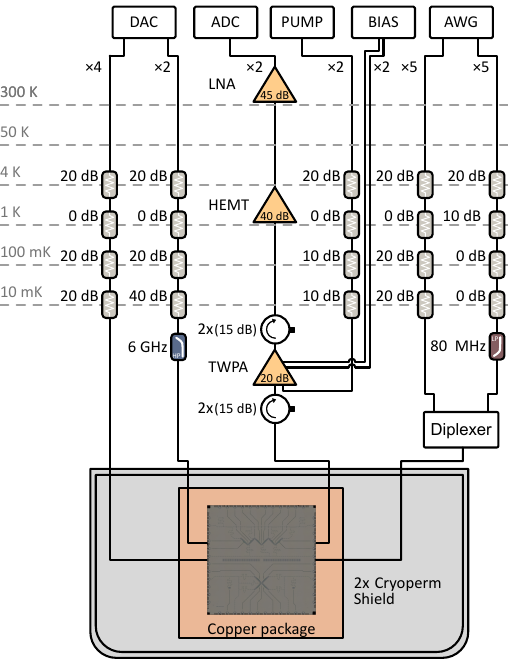}
    \caption{\textbf{Experimental setup.}}
    \label{fig:setup}
\end{figure}
The sample is mounted at the mixing chamber stage of a dilution refrigerator (\textit{Bluefors} XLD1000sl). 
Fig.~\ref{fig:setup} displays the full electronic setup up to the room temperature control.
We employ two devices for qubit control:
A \textit{Zurich Instruments} quantum controller (SHFQC) is used for driving the coupler as well as readout of the qubit state, while two channels of a \textit{Zurich Instruments} arbitrary waveform generator (HDAWG) are utilized for DC-flux biasing and AC-flux control.
The signal input is attenuated with -60\,dB  for the coupler drive and with -80\,dB for the readout tone.
The DC and AC signals are combined at the mixing chamber using a diplexer (\textit{Minicircuits} ZDPLX-2150-S+) with a DC-10\,MHz pass-band on low-pass channel and a 50-2000\,MHz pass-band on the high-pass channel.
Additionally, the DC line is filtered by an 80\,MHz low-pass filter (\textit{Minicircuits} VLFX80+) to increase isolation in the stop-band.
\begin{figure*}
    \includegraphics[]{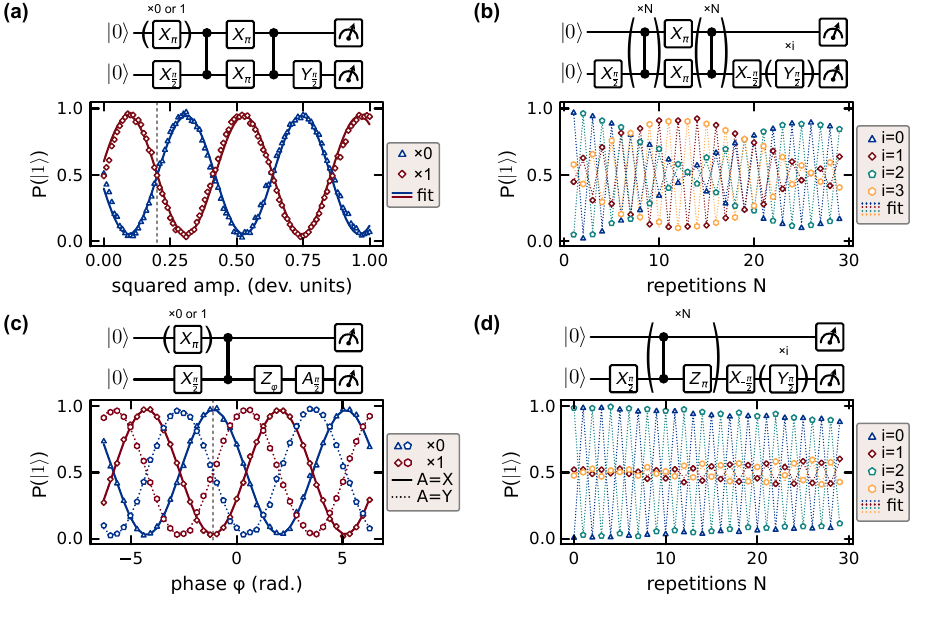}
    \caption{\textbf{Tune-up sequence for RIP gates.} (a) Amplitude calibration using a refocusing sequence. The grey dashed line indicates the optimal amplitude(b) Error amplification for amplitude errors using a repeated refocusing sequence. (c) Calibration of single-qubit virtual-$Z$ rotations with a Ramsey sequence. The grey dashed line indicates the optimal frame change. (d) Error amplification sequence for single-qubit virtual-$Z$ rotations.}
    \label{fig:tune-up}
\end{figure*}
The SHFQC generates and digitizes the readout signal, which is amplified by a TWPA (\textit{Arctic Instruments} AI-TWPA-C), a HEMT (\textit{LNF}-LNC4\_8C) and a room temperature amplifier (\textit{Qotana} DBLNA104000800F).
We pump the TWPA using a high-frequency microwave source (\textit{Anapico} APMS30G) and apply a DC bias using a floating DC source (\textit{Stanford Research Systems} SIM298).
We use a high-pass filter (\textit{Minicircuits} VHF-5050+) at the input port of the readout line in combination with two dual-junction isolators (\textit{LNF}-ISISC4\_12A) at the output to protect the qubits from unwanted noise photons in resonators.
The chip is packaged and mounted inside two cryoperm shields. 
We use on-device discrimination and conditional measurement feedback, which initializes each qubit in $\ket 0$ prior to each circuit execution.
\section{\label{appendix:tune-up}RIP-gate Tune-Up}
In the following, we describe the tune-up sequence used to calibrate the RIP gate. The sequences are agnostic to the specific gate implementation and apply to both the regular and the refocused RIP gate.
The propagator of a RIP gate, up to a global phase, is given by
\begin{equation}
    \hat U_\mathrm{RIP} = \begin{pmatrix}
        1 & 0 & 0 & 0\\
        0 & e^{i\alpha_\mathrm{A}} & 0 & 0\\
        0 & 0 & e^{i\alpha_\mathrm{B}} & 0\\
        0 & 0 & 0 & e^{i(\alpha_\mathrm{A}+\alpha_\mathrm{B}+\phi)}
    \end{pmatrix},
\end{equation}
with the single-qubit phases $\alpha_\mathrm{A}$ and $\alpha_\mathrm{B}$ and the conditional phase $\phi$. In the following, we discuss calibration methods for these three parameters in turn.

We first calibrate the conditional phase $\phi$ at an arbitrary fixed detuning $\Delta$ from $\omega^r_{00}$.
A discussion regarding the choice of the detuning can be found near the end of this section.
We record a refocused RIP sequence~\cite{paikExperimentalDemonstrationResonatorInduced2016, liRealizationHighFidelityCZ2024} shown in Fig.~\ref{fig:tune-up}\,(a), sweeping the drive amplitude $\epsilon$ applied to the coupler. Due to the echo, the sequence is insensitive to $\alpha_\mathrm{A}$ and $\alpha_\mathrm{B}$ and therefore probes only $\phi$. As the conditional phase accumulation scales with $|\epsilon|^2$~\cite{crossOptimizedPulseShapes2015}, we sample the amplitude with a quadratic spacing to obtain a linear oscillation. We repeat this measurement for both states of the control qubit and extract the drive amplitude that yields $\phi=\pi$ from a joint fit.

In the next step, we use error amplification~\cite{sheldonCharacterizingErrorsQubit2016, sungRealizationHighFidelityCZ2021, lazarCalibrationDriveNonlinearity2023} to fine-tune the pulse amplitude. As shown in Fig.~\ref{fig:tune-up}\,(b), we repeat the RIP gate $N$ times before and after the echo, amplifying over-rotation errors of $\phi$, followed by $i$ $Y_{\pi/2}$-rotations into the four cardinal directions of the target qubit. We record this sequence for up to 30 repetitions and fit the data using
\begin{equation}
    P_{\ket 1}(i, N) = \frac{A}{2}\left[1 - \cos\left(\frac{\pi}{2}i + N\pi\varepsilon\right)e^{-\gamma N}\right] + B,
    \label{eq:overrotation}
\end{equation}
where $\varepsilon$ is the rotation parameter with $\varepsilon=1$ for an ideal gate, $A$ a scale, $B$ an offset, and $\gamma$ a decay constant accounting for decoherence-induced contrast loss. The drive amplitude is iteratively updated until $|\varepsilon - 1|$ falls within the fit uncertainty. We record four cardinal directions rather than one as this stabilizes the fit over a broad range of $\varepsilon$, whereas the uncertainty of a single-direction fit diverges as $\varepsilon$ approaches 1.

With $\phi$ calibrated, we calibrate the single-qubit phases $\alpha_\mathrm{A}$ and $\alpha_\mathrm{B}$ in the same two-step manner—a coarse calibration followed by error amplification. The coarse step, shown in Fig.~\ref{fig:tune-up}\,(c), prepares the target qubit in a superposition state, followed by a single RIP gate, a virtual phase rotation, and a final $\pi/2$-pulse. By sweeping the virtual rotation phase and repeating the sequence with $X_{\pi/2}$ and $Y_{\pi/2}$ as the final rotation, and for both states of the control qubit, we obtain a virtual oscillation described by
\begin{equation}
    P_{\ket 1}(\varphi, \theta, i) = \frac{A}{2}\left[1 - \cos\left(\varphi + \theta + \frac{\pi}{2}i\right)\right] + B.
\end{equation}
Here, $i\in\{0,1\}$ selects the axis of the final $\pi/2$-pulse, $\varphi$ is the virtual rotation angle, and $\theta$ is the phase added by the RIP gate, taking the value $\theta_0 = \alpha_\mathrm{T}$ when the control qubit is in $\ket 0$ and $\theta_1 = \alpha_\mathrm{T} + \phi$ when it is in $\ket 1$, with $\mathrm{T}\in\{q_i,q_j\}$ denoting the target qubit. The sequence is repeated with qubit $q_i$ and qubit $q_j$ alternating in the roles of control and target.

Finally, we apply an error-amplification sequence, shown in Fig.~\ref{fig:tune-up}\,(d), to both $\alpha_\mathrm{A}$ and $\alpha_\mathrm{B}$. In analogy to amplification sequences used for phase shifts in single-qubit gates~\cite{luceroReducedPhaseError2010, lazarCalibrationDriveNonlinearity2023}, the phase error is amplified by repeated applications of the target gate followed by a virtual $Z_\pi$ rotation. Analogously to the calibration of $\phi$, we record four cardinal directions, fit the data using Eq.~(\ref{eq:overrotation}), and iterate until the residual phase error is within the fit uncertainty.

\begin{figure}
    \centering
    \includegraphics{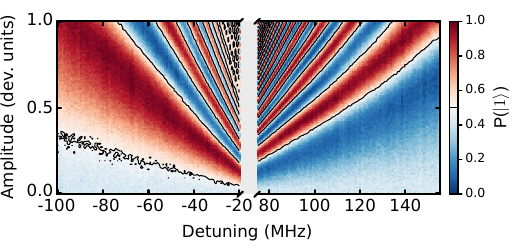}
    \caption{\textbf{Conditional phase vs drive detuning.} The qubit population encoding the conditional phase is recorded using a refocusing sequence as in Fig.~\ref{fig:tune-up}\,(a) for different $\Delta$ from $\omega^r_{00}$ for the coupler $r_{45}$.
    Note that the phase accumulation is opposite depending on positive or negative $\Delta$.}
    \label{figure7}
\end{figure}
Our calibration routine does not address the choice of detuning $\Delta$ to the resonator.
While theoretical treatment of the RIP gate suggests maximizing $\Delta$ is always optimal regarding photon-induced dephasing~\cite{crossOptimizedPulseShapes2015},
we find that in practice, $\mathcal{F}_\mathrm{CZ}$ only qualitatively follows this trend.
The occurrence of spurious resonances, which we assume to originate either from spurious resonances within the circuit or two-level fluctuators, can distort the acquired conditional phase or cause leakage, making the choice of $\Delta$ non-trivial~\cite{malekakhlaghOptimizationResonatorinducedPhase2022b}.
However, as exemplified in Fig.~\ref{figure7}, the conditional phase can be set within a wide range of detunings, such that circumventing such spurious resonances is always possible.
In parts, the small detuning between all coupler frequencies $\omega_{r_{ij}}$ stated in Table~\ref{tab:coupler-parameters} is responsible for a more challenging determination of the best $\Delta$ as well.
\section{\label{appendix:qst}Quantum state tomography}
We record the presented state-fidelities using QST in the following manner.
First, we apply the circuit preparing the state. 
We then use single-qubit rotations to prepare all $3^N$ combinations of Pauli-correlators $\mathcal P\in\{X,Y,Z\}^{\otimes N}$, followed by simultaneous readout of all qubits.
Combinations involving the identity $I$ are extracted from partial correlators of this set.
Since the median readout errors of the device are significant (median readout fidelity of $94_{-3}^{+1}$\,\% with the smallest measured fidelity at 85.5\,\%), we additionally prepare and measure each computational state to record the detector error matrix defined by $\mathbf p_\mathrm{exp}=M\mathbf p_\mathrm{ideal}$.
To ensure that potential fluctuations in time affect $M$ and the QST equally, we record both experiments with interleaved single-shots.
We obtain the Pauli expectation values from the corrected outcome probabilities averaged over redundant terms, yielding a linear-inversion estimate of the density matrix.
This step is carried out once for the probabilities with and without measurement error correction.
To ensure a physical density matrix with $\mathrm{tr}\rho=1$ and $0\leq\rho_{ij}\leq1$ for the probabilities corrected for measurement errors, we refine the estimate of $\rho$ by maximum likelihood estimation.
We obtain error-corrected state-fidelities of 96.9(2)\,\%, 95.4(2)\,\% and $90.0(2)$\,\% for GHZ-states involving 3,4 and 5 qubits as well as raw fidelities of 64.8(1)\,\%, 68.5(2)\,\% and 62.4(1)\,\% respectively. 
We attribute the comparatively low raw fidelity of the 3-qubit GHZ-state to a dropout in readout performance visible in $M$.
All uncertainties reported for the QST are determined by bootstrapping the dataset and running the analysis over 50 equally sized subsets of the data.

\bibliography{bibliography.bib}
\end{document}